\documentclass[12pt]{article}

\begin{document} 

\vspace*{1.1cm} 
  \begin{center} 
  {\Large \bf Local contribution of a quantum condensate to the vacuum 
energy density} 
  \end{center} 

  \begin{center} 
  \vskip 10pt 
  Giovanni Modanese\footnote{e-mail address: 
  giovanni.modanese@unibz.it}
  \vskip 5pt
  {\it California Institute for Physics and Astrophysics \\
  366 Cambridge Ave., Palo Alto, CA 94306}
  \vskip 5pt
  and
  \vskip 5pt
  {\it University of Bolzano -- Industrial Engineering \\
  Via Sernesi 1, 39100 Bolzano, Italy}
  
\vskip 10pt

\end{center} 

\baselineskip=.175in 
    
\begin{abstract}
We evaluate the local contribution $g_{\mu \nu}L$ of coherent 
matter with lagrangian density $L$ to the vacuum energy density. Focusing on the case of 
superconductors obeying the Ginzburg-Landau equation, we express the 
relativistic invariant density $L$ in terms of low-energy quantities containing 
the pairs density. We discuss under which physical conditions the 
sign of the local contribution of the collective wave function to the vacuum 
energy density is positive or negative. Effects of this kind can play an 
important role in bringing about local changes in the amplitude of 
gravitational vacuum fluctuations - a phenomenon reminiscent of the Casimir effect in QED.
\end{abstract}

\bigskip

\noindent
{\bf 1. Global cosmological term}
\medskip

The vacuum energy density term, or ``cosmological term" in the Einstein equations 
has become popular again in the last years, after a series of 
observations which indicate $\Lambda \sim 10^{-50} \ 
cm^{-2}$ as most probable value. Earlier observations set an upper limit on $\Lambda$ of the order of $10^{-54} 
\ cm^{-2}$, so it was thought to vanish exactly for symmetry 
reasons. 

As remarked by several authors, the observed non-zero value of $\Lambda$ creates an arduous fine-tuning problem \cite{RPP}. 
This value should in fact be regarded as the residual of a complex interplay, 
in poorly known sectors of particle physics, between positive and negative 
vacuum energy densities. According to \cite{Shapiro}, the observed residual 
should also be scale-dependent and this dependence could appear most 
clearly at length scales corresponding to the mass of the lightest 
particles, like neutrinos or unidentified scalars.

The above mentioned value of $\Lambda$ is global, in the sense that it is supposed to define a uniform background present in the whole universe.
In order to elucidate the notion of local contributions to the vacuum energy density, let us first recall here some known facts. The cosmological term is an addition to the l.h.s.\ of the Einstein equations, of the form $\Lambda g_{\mu \nu}(x)$, with $\Lambda$ constant:
\begin{equation}
R_{\mu \nu}(x) - \frac{1}{2} g_{\mu \nu}(x) R(x) + \Lambda g_{\mu \nu}(x) = -8\pi G T_{\mu \nu}(x).
\end{equation}

A quantum field $\Phi$ contributes to the cosmological constant in two ways:

(i) Its zero-point oscillations are associated with a huge energy density, whose value depends on the frequency cut-off.
 
(ii) The field can have a non-vanishing vacuum expectation value (VEV) $\Phi_0= \langle 0|\Phi|0 \rangle$. This contribution is essentially classical. If $L$ is the field lagrangian, the corresponding cosmological term is $-8\pi GL(\Phi_0)g_{\mu \nu}(x)$, because the energy-momentum tensor has in general the form 
\begin{equation}
T_{\mu \nu}=\Pi_\mu \partial_\nu \Phi - g_{\mu \nu}L,
\end{equation}
where $\Pi_\mu$ (conjugate momentum of $\Phi$) and $\partial_\nu \Phi$ vanish in the case of a constant field. One of the roots of the cosmological constant paradox is clearly seen here: any constant term in the lagrangian of a system, though irrelevant for the system's dynamics, has gravitational consequences. In elementary particle physics a non-vanishing VEV is usually the consequence of a spontaneous symmetry breaking process.

\medskip
\noindent
{\bf 2. Local cosmological term}
\medskip

A {\it local} contribution to the vacuum energy density can arise when the state of a localized physical system is described by a classical field comparable with the VEV of a quantum field. We are interested into cases of this kind occurring in condensed matter physics. In this context, the physical systems properly described by {\it continuous} classical-like fields (also at microscopic level, not just in a macroscopic-average sense as for fluids) are basically 

(1) the electromagnetic field in the low-frequency limit, in states where the photons number uncertainty is much larger than the phase uncertainty; 

(2) systems with macroscopic quantum coherence, described by ``order parameters", like superfluids, superconductors and spin systems.

Suppose that, in one of these systems, the field has a constant value $\Phi_0$ in a bounded region and is zero outside. (Consider for instance a container with superfluid helium of constant density.) We can speak of a contribution of the field to the cosmological constant in this region, equal to $-8\pi GL(\Phi_0)$. A magnitude order estimate shows that such a contribution can be of the same order of the observed $\Lambda$ or even larger, although it is clear that this energy, being present only in a small region of space, does not have any cosmological relevance. 

From the classical point of view, one can correctly object that the description of this situation in terms of a local cosmological constant is purely formal, because the gravitational field present is just that due to the superfluid regarded as an energy-momentum source. Moreover, there is no distinction, still at the classical level, between a truly continuous source, like the superfluid wave function, and an incoherent fluid. 

\medskip
\noindent
{\bf 3. The cosmological term vs.\ dipolar fluctuations}
\medskip

The perspective changes if one takes into account short-scale gravitational quantum fluctuations. Suppose to describe gravity with the covariant perturbation theory in the 
weak-field approximation. The action contains some parameters, and one of these is the
effective $\Lambda$ in the considered region. The value of $\Lambda$ acquires a special significance in relation to the so-called ``dipolar vacuum fluctuations", which constitute a class of zero modes of the pure Einstein action \cite{vdf}. Let us recall their features in short.

 The functional integral of pure Einstein quantum gravity 
 can be written as $z = \int d [g_{\mu \nu}] 
 \exp(i S/ \hbar)$, with 
 $S = \int d^4x \sqrt{g(x)} R (x)/8\pi G$. 
 The ``spacetime foam" \cite{foam} consists of fluctuations 
 whose action does not exceed a quantity of order $\hbar$.
\footnote{Note, however, that we use units such that $\hbar=c=1$ throughout this work.} 
 This implies, for curvature fluctuations on a scale $r$, that   
 $| R | < G/r^4$ 
 (according to naive power counting) or 
 $| R | < 1/(L_{P}r)$  
 (according to numerical lattice estimates \cite{hw}). 
 Therefore, large fluctuations are expected to take place 
 only at very small distances. 
  
 Since, however, the Einstein action is not positive
 definite, one can also expect some fluctuations due to 
 peculiar cancellations of distinct contributions to the 
 action, which are by themselves larger than $\hbar$. 
 In order to show this explicitely, let us consider the 
 Einstein equations with an 
 {\it auxiliary} source $T_{\mu \nu}$: 
  \begin{equation} 
 R_{\mu \nu}(x) - \frac{1}{2} 
 g_ {\mu \nu}(x) R(x) = 
 -8 \pi G T_{\mu \nu}(x) 
 \label{uno} 
 \end{equation} 
  and their trace 
  \begin{equation} 
 R(x) = 8 \pi G g^{\mu \nu}(x) 
 T_{\mu \nu}(x) \equiv 8 \pi G 
 {\rm Tr} T (x) 
 \label{due} 
 \end{equation} 
  
 Then consider a solution $g_{\mu \nu}(x)$ of (\ref{uno}) 
 with any source satisfying the condition 
  \begin{equation} 
 \int d^4x \sqrt{g(x)} {\rm Tr} 
 T(x) = 0 
 \label{croce} 
 \end{equation} 
  In view of (\ref{due}), this metric has zero action. We 
 have constructed in this way a zero mode of the pure 
 Einstein action. The source is unphysical, but it is 
 ``forgotten" after obtaining the metric. Condition 
 (\ref{croce}) means in fact that it is a ``dipolar" source, 
 with a compensation between regions having positive and 
 negative mass-energy density. Since this auxiliary source 
 is used to construct a virtual field configuration, we 
 sometimes call it a ``virtual source". 
  
In the presence of a cosmological term in the action, the dipolar fluctuations acquire a small monopolar component. In fact, the contribution of a dipolar fluctuation to the cosmological term is of the form \cite{vdf}
\begin{equation}
	\Delta S_\Lambda = \Lambda \tau M r^2 Q,
\label{deltas}
\end{equation}
where $\tau$ is the duration of the fluctuation, $M$ is the order of 
magnitude of the virtual +/- masses, $r$ their distance and $Q$ is an 
adimensional function which can be positive or negative and depends on the detailed form of the dipole. 
This means that it is still possible to obtain a zero mode, provided the contribution to the Einstein part of the action is the opposite of $\Delta S_\Lambda$. This implies in turn that the integral of the covariant trace of the energy-momentum tensor needs not be zero, i.e., there is a monopolar mass component in the virtual source. 

The magnitude order of the monopolar component is readily found to be 
\begin{equation}
|m_{tot}| \sim \Lambda M r^2
\end{equation}
Here the amplitude $M$ of the dipolar component is limited only by the $R^2$-like quantum corrections to the Einstein action (Ref.\ \cite{vdf}, Section 2.4). Choosing length and time scales as $r \sim 1~cm$, 
$\tau \sim 1~s$ ($\sim 10^{10}~cm$ in natural units), one finds an upper limit for $M$ of $\sim 10^{56}~cm^{-1}$; with the observed value of $\Lambda$, the monopolar residual is then $|m_{tot}| \sim 10^{6}~cm^{-1}$.

So we see that the effective local value of the cosmological constant sets the amplitude of the monopolar component of a class of strong quantum gravity fluctuations, corresponding to remarkable positive and negative virtual mass densities, and present at distance scales much larger than Planck scale. 
\footnote{In a complete, non-perturbative theory of quantum gravity these fluctuations should of course take part in defining the renormalized value of the coupling constants and of quantities like the vacuum polarizability.}
Consider now the local contribution of coherent matter to the cosmological term again . If this causes some local inhomogeneity in $\Lambda$, an inhomogeneity in the vacuum fluctuations will follow, and it is well known from the Casimir effect that this can give rise to observable effects. In an analogy with the Casimir effect, the role of the metallic conductors which impose a cut-off on the electromagnetic vacuum fluctuations is played here by the coherent condensed matter.

\medskip
\noindent
{\bf 4. General form of $L$ for a superconductor obeying the Ginzburg-Landau equation, in the non-relativistic limit}
\medskip

At this point it is important to evaluate the lagrangian density $L$ in some concrete situation and compare its value, in sign and magnitude, with the background $\Lambda/8\pi G$. For the electromagnetic field this is straightforward. Superfluids, on the other hand, cannot be adequately described by an effective lagrangian theory \cite{Tilley}. So we shall focus our attention on the case of superconductors, for which a widely used lagrangian
phenomenological theory is available -- the Ginzburg-Landau theory -- based 
on the macroscopic wave function $\psi_{GL}$. In the following we shall compute the relativistic lagrangian of a superconductor in terms of the non relativistic wave 
function $\psi_{GL}$, which is well known in several physical situations. 

We are therefore seeking the low energy limit of a scalar field theory appropriate for superconductors. This is not trivial.
Some proposals for the inverse procedure, namely a relativistic 
generalization of Ginzburg-Landau theory, have been previously described in 
the literature \cite{Anandan,Russi}. However, in \cite{Anandan} only 
variations in the phase of $\psi_{GL}$ are considered, and in \cite{Russi} 
the tetrad formalism is employed, with a particular gauge fixing.
We shall follow a somewhat more conservative method. Let us consider the 
usual time-independent non-relativistic limit of the Klein-Gordon wave 
function, namely 
\begin{equation}
	\psi_{KG}({\bf x})= e^{imt} \phi(x),
\end{equation}
where $m$ is the Cooper pairs mass.
By introducing this into the Klein-Gordon lagrangian one obtains
\begin{equation}
	L  = - \frac{1}{2} \nabla \psi_{KG}^*({\bf x}) \nabla \psi_{KG}({\bf 
x}) .
\end{equation}
This coincides with the part of the Ginzburg-Landau lagrangian containing 
the partial derivatives \cite{Waldram}, provided the wave function is 
suitably normalized. We set
\begin{equation}
	\psi_{GL}({\bf x})= \sqrt{m} \psi_{KG}({\bf x})
\end{equation}
and so obtain
\begin{equation}
	L  = - \frac{1}{2m} \nabla \psi_{GL}^*({\bf x}) \nabla \psi_{GL}({\bf 
x}).
\end{equation}
In the following we shall just write $\psi$ instead of $\psi_{GL}$ and we 
shall omit the space dependence. The normalization above corresponds to 
$|\psi({\bf x})| =\rho({\bf x})$, where $\rho^2$ is the density of Cooper 
pairs.

Next we generalize our initial lagrangian, without spoiling its 
relativistic invariance, in such a way to make it coincide with the full 
Ginzburg-Landau lagrangian in the low energy limit. To this end it suffices 
to add a quadratic and a quartic term. We also introduce the minimal coupling 
with the four-potential ${\bf A}({\bf x})$, finally obtaining 
\begin{equation}
	L  = - \frac{1}{2m} | -i \nabla \psi + 2e{\bf A}\psi|^2 - \alpha 
\psi^* \psi - \frac{1}{2} \beta (\psi^* \psi)^2=
\end{equation}
\begin{equation}
	= - \frac{1}{2m} \left[ (\nabla \psi^*) (\nabla \psi) - 2i {\bf 
A}\psi^* \nabla \psi + 2i {\bf A}\psi \nabla \psi^* + 4e^2 {\bf A}^2 \psi^* 
\psi \right]
- \alpha \psi^* \psi - \frac{1}{2} \beta (\psi^* \psi)^2,
\label{zero}
\end{equation}
where $\alpha$ and $\beta$ are two arbitrary coefficients, which in the end 
are identified with the Ginzburg-Landau coefficients. This 
lagrangian density, multiplied by $-8\pi G$, gives the local 
contribution to the vacuum energy density. Now we evaluate it for wave 
functions which satisfy the Ginzburg-Landau equation. 

A crucial point should be stressed here. Remember that this 
equation is obtained by minimizing the spatial {\it integral} of $L$; while 
doing this, one transforms a term $(\nabla \psi^*) (\nabla \psi)$ into a 
term $\psi^* \nabla^2 \psi$, integrating by parts and supposing that 
$\psi=0$ at the boundary. Here, however, we are not interested into the 
integral of $L$, but into its local value. Therefore, we take the 
Ginzburg-Landau equation 
\begin{equation}
	\frac{1}{2m} \left( -i \nabla \psi + 2e {\bf A} \right)^2 \psi + 
\left(\alpha + \beta \psi^* \psi \right) \psi = 0
\label{gl}
\end{equation}
and multiply it by $\psi^*$ on the left. In the gauge $\nabla {\bf A}=0$, 
$\nabla$ commutes with ${\bf A}$; after isolating the term $4e^2 {\bf A}^2 
\psi^* \psi$ and replacing it into (\ref{zero}), we find
\begin{equation}
	L = - \frac{1}{2m} \left[ (\nabla \psi^*) (\nabla \psi) - 2ie {\bf A} (\psi^* \nabla \psi + \psi \nabla \psi^*) +  \psi^* \nabla^2 \psi -m\beta 
(\psi^* \psi)^2  \right].
\label{quasifinale}
\end{equation}

We observe that $L$ is, by construction, real. Therefore the imaginary part 
of the expression above must vanish and we have
\begin{equation}
	2ie{\bf A} (\psi^* \nabla \psi + \psi \nabla \psi^*)=i Im (\psi^* 
\nabla^2 \psi).
\end{equation}
In conclusion equation (\ref{quasifinale}) becomes
\begin{equation}
	L = - \frac{1}{2m} \left[ (\nabla \psi^*) (\nabla \psi) +  Re(\psi^* 
\nabla^2 \psi) - m\beta (\psi^* \psi)^2  \right].
\label{finale}
\end{equation}

\medskip
\noindent
{\bf 5. Numerical evaluation of $L$}
\medskip

Introducing variables $\rho({\bf x})$ and $\theta({\bf x})$, such that 
$\psi({\bf x})=\rho({\bf x}) e^{i\theta({\bf x})}$, we finally obtain for the lagrangian density a 
simple expression, in which the magnetic field does not appear explicitly
\begin{equation}
	L = - \frac{1}{2m} \left[ (\nabla \rho)^2 + \rho \nabla^2 \rho -m 
\beta \rho^4  \right].
\label{finaleconrho}
\end{equation}
In this expression we can restore $\hbar$, obtaining
\begin{equation}
	L = - \frac{1}{2m} \left[ \hbar^2(\nabla \rho)^2 + \hbar^2 \rho 
\nabla^2 \rho -m \beta \rho^4  \right].
\label{finaleconrho2}
\end{equation}

We recall \cite{Waldram} that the $\alpha$ and $\beta$ coefficients depend 
on the absolute temperature $T$. The coefficient $\beta$ is always positive 
and approximately constant near $T_c$; $\alpha$ is negative for $T<T_c$ and 
behaves like $\sim const. (T-T_c)$ near $T_c$. The ratio between $\alpha$ and 
$\beta$ is given by the relation $n_p=-\alpha / \beta$, where $n_p$ is the 
average density of pairs in the material. Finally, $\beta$ is linked to the 
value of the Ginzburg-Landau parameter $\kappa=\lambda/\xi$ by the relation 
$\kappa^2 = m^2 \beta/(2\mu_0 \hbar^2 e^2)$. We also recall that the wave 
function must satisfy suitable boundary conditions at interfaces with 
vacuum and at junctions with normal conductors. 

It is straightforward to check that the sign of $L$ is positive for two types of configurations:

(1)	For the constant solutions of eq.\ (\ref{gl}) in the absence of 
external field, which implies $\rho^2({\bf x})=n_p$. The corresponding 
constant lagrangian density is $L_1=\frac{1}{2} \beta n_p^2$.

(2)	For regions of the superconductor where $\rho \nabla^2 \rho$ is negative 
and greater, in absolute value, than $(\nabla \rho)^2$. It is 
straightforward to check that these are regions located around local 
density maximums, or more generally lines and surfaces where the 
first partial derivatives of $\rho$ are zero and the second 
derivatives are negative or null. The lagrangian density at a maximum 
is $L_2 \sim \frac{\hbar^2}{2m} \rho |\rho ''|$. If the maximum is sharp, $L_2$ 
can be much larger than $L_1$. Configurations of 
this kind are characteristic of solutions of the Ginzburg-Landau 
equation with strong magnetic flux penetration \cite{Tilley}.

In all other configurations, $L$ is negative. Some numerical estimates are given in Table 1, where the gradients are taken to be of the order of $\rho \xi^{-1}$. At local minima or in regions with strong gradients, we can suppose that $|L|$ is of the same magnitude order as $L_2$.

\begin{table}
\caption{Magnitude orders of the lagrangian densities $L_1$ and $L_2$ for a type I superconductor (Pb) and for a type II superconductor (YBCO), computed according to eq.\ (\ref{finaleconrho2}). Also listed are the values of the London length $\lambda$ and the coherence length $\xi$ at $T=0$ \cite{Waldram}and the average pairs density $n_p$, computed from $\lambda$ by the relation $n_p=m_e/(2\mu_0 e^2 \lambda^2)$. For YBCO, the values of $\xi$ along the $a$-$b$ direction and the $c$ direction are given separately, and so the corresponding values of $L_2$; $n_p$ is the density in the $a$-$b$ planes, computed with the effective mass $m^* \simeq 4.5 m_e$.}

\medskip

\begin{tabular}{|c|c|c|}
\hline
& {\bf Pb} & {\bf YBCO} \\
\hline
$\lambda$ ($m$) & $3.9 \cdot 10^{-8}$ & $1.4 \cdot 10^{-7}$ ($a$-$b$ direction)\\
$\xi$ ($m$) & $8.2 \cdot 10^{-8}$ & $1.6 \cdot 10^{-9}$ ($a$-$b$ direction) \\
& & $2.4 \cdot 10^{-10}$ ($c$ direction) \\
$n_p$ ($m^{-3}$) & $9.3 \cdot 10^{27}$ & $3.5 \cdot 10^{27}$ \\
$L_1$ ($Jm^{-3}$) & $10^{4}$ & $10^{6}$ \\
$L_2$ ($Jm^{-3}$) & $10^{4}$ & $10^{6}$ ($a$-$b$ direction) \\
& & $10^{8}$ ($c$ direction) \\
\hline
\end{tabular}
\end{table}

\medskip
\noindent
{\bf 6. Conclusions}
\medskip

We are interested into the value of the lagrangian density $L$ for condensed matter systems described by an order parameter, because the value of $L$ - by interfering with the vacuum energy term in the Einstein action - affects locally the amplitude of dipolar vacuum fluctuations. This amplitude is proportional to $|L-\Lambda/8\pi G|$, where $\Lambda$ is the ``natural" background vacuum energy density. At cosmological scale $\Lambda$ is known to be negative (with our conventions on the metric) and of the order of $10^{-50} \ cm^{-2}$, so that $|\Lambda/8\pi G|\sim 10^{-1}~J/m^3$ in SI units. However, $\Lambda$ is probably scale-dependent, and increases at small distances \cite{Shapiro}.

For electric and magnetic fields in the low frequency limit, when the photons number uncertainty is much larger than the phase uncertainty, the lagrangian density is simply proportional to the square of the field strength and has positive and negative sign, respectively. The magnitude order of $L$ is, for instance, $L_E \sim 10^2~J/m^3$ for $E= 10^6~V/m$ and $ L_B \sim -10^5~J/m^3$ for $B=1~T$.

For a superconductor described by a collective wave function obeying the Ginzburg-Landau equation, we have seen that in the non-relativistic limit it is possible to express $L$ as a function of $|\psi_{GL}|$. Relevant values of $L$ are those for the case of constant pairs density ($L_1$) and those at local density maximums ($L_2$). As shown in Table 1, this kind of local contributions to the vacuum energy density can be much larger than the pure electromagnetic contributions.

In the absence, however, of any reliable experimental or theoretical estimates for the value of $\Lambda$ at short scales, it is not possible to compare $\Lambda/8\pi G$ with the above values of $L$ and guess whether the lagrangian density of coherent matter can affect the amplitude of the monopolar component of dipolar vacuum fluctuations. Conversely, the experimental search for anomalous effects of the gravitational fluctuations in the presence of coherent matter can give useful information on the short-scale value of $\Lambda$ (see \cite{vigier} and ref.s).

The hypothesized phenomenon could be regarded as a sort of gravitational analogue of the Casimir effect. In spite of the relative weakness of gravitational forces at atomic scale, such vacuum effects can be relevant because the dipolar fluctuations in Einstein gravitation are zero-modes of the action and are therefore much stronger than electromagnetic vacuum fluctuations.

\bigskip \noindent {\bf Acknowledgment} - This work was
 supported in part by the California Institute for Physics
 and Astrophysics via grant CIPA-MG7099.

\end{document}